\documentclass{aa}
\usepackage{graphicx}
\usepackage{txfonts}
\usepackage{natbib}
\usepackage{url}

\begin{document}

   \title{Signatures of intrinsic Li depletion and Li-Na anti-correlation in the metal-poor globular cluster NGC~6397 \thanks{Based on data collected at European Southern Observatory (ESO), Paranal, Chile, under program IDs 077.A-0018(A) and 281.D-5028(A), as well as data collected with the Danish 1.54\,m at European Southern Observatory (ESO), La Silla.}\fnmsep\thanks{Tables 2, 3, and 4 are only available in electronic form at the CDS via anonymous ftp to cdsarc.u-strasbg.fr (130.79.128.5) or via http://cdsweb.u-strasbg.fr/.}}

   \author{K. Lind\inst{1}\and
           F. Primas\inst{1}\and
           C. Charbonnel\inst{2,3}\and
           F. Grundahl\inst{4}\and
           M. Asplund\inst{5}
           }

   \institute{European Southern Observatory (ESO), Karl-Schwarzschild-Strasse 2,
              857 48 Garching bei M\"unchen, Germany\\
              \email{klind@eso.org}\and
              Geneva Observatory, 51 chemin des Mailettes, 1290 Sauverny, Switzerland\and
              Laboratoire d'Astrophysique de Toulouse-Tarbes, CNRS UMR 5572, Universit\'e de Toulouse, 14, Av. E.Belin, F-31400 Toulouse, France\and
              Department of Physics \& Astronomy, Aarhus University, Ny Munkegade, 8000 Aarhus C, Denmark\and
              Max-Planck-Institut f\"ur Astrophysik, Karl-Schwarzschild-Strasse 1,
              857 41 Garching bei M\"unchen, Germany
             }
   \date{Received 2009 May 19; accepted 2009 June 10}

\authorrunning{K.Lind et al.}  \titlerunning{Intrinsic Li depletion and Li-Na anti-correlation in NGC~6397}

  \abstract
   {To alleviate the discrepancy between the prediction of the primordial lithium abundance in the universe and the abundances observed in Pop II dwarfs and subgiant stars, it has been suggested that the stars observable today have undergone photospheric depletion of lithium.}
   {To identify the cause of this depletion, it is important to accurately establish the behaviour of lithium abundance with effective temperature and evolutionary phase. Stars in globular clusters are ideal objects for such an abundance analysis, because relative stellar parameters can be determined precisely.}
   {We have conducted a homogeneous analysis of a very large sample of stars in the metal-poor globular cluster NGC~6397, covering all evolutionary phases from below the main sequence turn-off to high up on the red giant branch. Non-LTE Li abundances or abundance upper limits were obtained for all stars, and for a sizeable subset of the targets sodium abundances were also obtained. The Na abundances were used to distinguish stars formed out of pristine material from stars formed out of material affected by pollution from a previous generation of more massive stars.}
   {The dwarf, turn-off, and early subgiant stars in our sample form a thin abundance plateau, disrupted in the middle of the subgiant branch by the Li dilution caused by the first dredge-up. A second steep abundance drop is seen at the luminosity of the red giant branch bump. The turn-off stars are more Li-poor, by up to 0.1\,dex, than subgiants that have not yet undergone dredge-up. In addition, hotter dwarfs are slightly more Li-poor than cooler dwarfs, which may be a signature of the so-called Li dip in the cluster, commonly seen among Pop\,I stars. The feature is however weak. A considerably wide spread in Na abundance confirms that NGC~6397 has suffered from intracluster pollution in its infancy and a limited number of Na-enhanced and Li-deficient stars strongly contribute to forming a significant anti-correlation between the abundances of Na and Li. It is nevertheless seen that Li abundances are unaffected by relatively high degrees of pollution. Lithium abundance trends with effective temperature and stellar luminosity are compared to predictions from stellar structure models including atomic diffusion and ad-hoc turbulence below the convection zone. We confirm previous findings that some turbulence, with strict limits to its efficiency, is necessary for explaining the observations.}
   {}


   \maketitle

\section{Introduction}
The detection of Li in the atmospheres of old and metal-poor Population II stars has opened up an observational link to the primordial Universe. $^7$Li is indeed one of four isotopes that were synthesised by nuclear reactions shortly after the Big Bang. With the most recent determination of the baryon density from the 5-year release of WMAP data, $\Omega_bh^2=0.02273\pm0.00062$ \citep{Dunkley09}, an initial abundance of $N(\rm^7Li)/\it N(\rm H)=5.24^{+0.71}_{-0.67}\times10^{-10}$ or $A\rm(Li)=2.72\pm0.06$\footnote{$A\rm(Li)=\it\log{\left(\frac{N\rm(Li)}{N(\rm H)}\right)}\rm+12$} is obtained from standard Big Bang nucleosynthesis (BBNS) \citep{Cyburt08}. This is significantly higher than what is found for the ``Spite plateau" \citep{Spite82} in the Galactic halo, a well-defined Li abundance plateau consisting of metal-poor halo dwarfs and early subgiant branch (SGB) stars. 

The plateau abundance in the field has been determined in a number of recent studies to span the range $A\rm(Li)\approx2.0-2.4$ for stellar metallicities $\rm[Fe/H]=(-3.5)-(-1.0)$, with a possible tendency toward increasing Li abundance with increasing metallicity \citep[e.g.][]{Ryan01,Melendez04,Charbonnel05a,Asplund06,Bonifacio07a,Hosford09,Aoki09}. Differences between the various analyses may arise from the choice of effective temperature scale and corrections for non-LTE effects. It is nevertheless clear that a discrepancy by a factor of $2-4$ with the cosmological prediction is unavoidable, suggesting that the stars have undergone the corresponding surface depletion of Li. An important fact is that in all the recent observational analysis (i.e., 21st century) of halo field stars, no evidence of significant dispersion has been found along the plateau, except for a few stars with strong Li abnormalities \citep[e.g.][]{Ryan99,Asplund06}. 

Soon after the discovery of the Spite plateau, \citet{Michaud84} predicted Li depletion under the combined action of gravitational settling and weak turbulence in the radiative zones of Population II stars. Using sophisticated stellar models that treat atomic diffusion and radiative acceleration from first principles, \citet{Richard05a} illustrate how the comparably low Li value found in old, metal-poor stars could be naturally explained, assuming rather strict limits to the turbulent transport extent and efficiency. However, as of today the nature of the underlying physical mechanism responsible for turbulence has not been definitively identified, the main difficulty for the conjectured processes (mass loss, rotation-induced mixing, ...) being to account for the negligible Li dispersion along the plateau.

\citet{Charbonnel05a} revisited the literature Li data for halo field stars with particular focus on the evolutionary status of the sample stars. This study discovered for the first time that the mean Li value appear to be higher for the turn-off (TO) and SGB stars than for the dwarfs. This result, together with the finding that all halo stars with Li abnormalities (i.e., strong deficiency or high content) lie on, or originate from, the hot side of the plateau, lead the authors to suggest that the most massive (i.e., post-MS) of the halo stars still observable today have had a different Li history than their less massive dwarf counterparts. \citeauthor{Charbonnel05a} suggested that such a behaviour may be the signature of a transport process of chemical elements and angular momentum whose efficiency changes on the blue edge of the plateau. This behaviour corresponds to that of the generation and filtering of internal gravity waves in both Population II and I stars \citep{Talon03,Charbonnel08}. \citet{Talon04} describe how internal gravity waves coupled with rotation-induced mixing are expected to lead to higher Li homogeneity among the plateau dwarf stars than among the more massive, slightly evolved stars, a scenario which may explain the observational findings by Charbonnel \& Primas. Speaking in favour of this model is its ability to simultaneously explain the internal solar rotation profile and the time evolution of the Li abundance at the surface of solar- and F-type stars as seen in Galactic open clusters \citep{Charbonnel05c}.

By turning to stars in metal-poor Galactic globular clusters, the intrinsic stellar processes involved can be further constrained. This approach presents some obvious advantages. In particular, the evolutionary status of the observed stars can be determined unambiguously and, presumably, all stars in a cluster were born with the same metallicity (although the surface metallicities observed today may vary between stars in different evolutionary phases, due to the effects of atomic diffusion \citep[e.g.][]{Korn07,Lind08}). 

NGC~6397 is one of the most well-studied metal-poor globular clusters and its Li content has been documented down to the magnitude of the cluster TO point in several studies \citep{Pasquini96,Castilho00,Thevenin01,Bonifacio02a,Korn06a,Korn07}, however, with rather poor number statistics. Bonifacio et al.\ found $A\rm(Li)=2.34\pm0.06$ for the mean abundance of twelve TO stars, whereas Korn et al.\ found $2.24\pm0.05$ for five TO stars and $2.36\pm0.05$ for two SGB stars. By comparing this observed abundance difference of Li (as well as Fe, Ca, Mg, and Ti) to predictions from the stellar-structure models by \citet{Richard05a} and references therein, Korn et al.\ empirically constrain the efficiency of slow macroscopic motions counteracting atomic diffusion below the convective envelope, which should ultimately provide clues on the origin of turbulence. Note that some slow macroscopic process is also required to reproduce the observed morphologies of globular cluster colour-magnitude diagrams \citep{Vandenberg02}.

Interpreting abundance trends in globular clusters must however be done with great caution. Indeed, it is well-known that globular-cluster stars present striking anomalies in their light element content that are not seen among their field counterparts \citep[for reviews see e.g.][]{Gratton04,Charbonnel05b}. More precisely, C, N, O, Na, Mg, and Al abundances show large star-to-star variations within individual clusters. C and N, O and Na,  and Mg and Al are respectively anti-correlated, the abundances of C, O, and Mg being depleted in some stars while those of N, Na, and Al are enhanced. Importantly, the abundance of Li was found anti-correlated with that of Na and correlated with that of O in turnoff stars in the more metal-rich globular clusters NGC~6752 and 47~Tuc \citep{Pasquini05,Bonifacio07b}. These abundance patterns are explained by contamination of the star-forming gas by hydrogen-processed material ejected by a first generation of short-lived massive globular-cluster stars. This increases the Na abundances while lowering the O and Li abundances with respect to the pristine cluster composition in a second generation long-lived low-mass stars. In the framework of the self-enrichment scenario the Li content of globular-cluster stars is actually an important tool to quantify the dilution factor between the ejecta of the massive stars responsible for pollution and the pristine intra-cluster matter \citep{Prantzos06,Decressin07a,Decressin07b}. \citet{Gratton01} and \citet{Carretta05} uncovered rather large variations in both O and Na, anti-correlated with each other, in SGB and dwarf stars in NGC~6397 \citep[see also][for bright giants]{Norris95,Castilho00}. This cluster has thus suffered from internal pollution in its infancy.

In the present study we analyse Li and Na abundances for a large sample of stars in NGC~6397, in an attempt to disentangle the primordial value of Li, with effects of atomic diffusion and intrinsic stellar depletion on the one hand, and early cluster pollution on the other hand. \S~2 describes the observations and data reduction, \S~3 describes the determination of stellar parameters and the abundance analysis. In \S~4 we present the Li and Na abundances found, and in \S~5 we discuss signatures of Li depletion in the cluster. \S~6 summarises our conclusions. 

\section{Observations}

The observations include spectroscopic data, described in \S~$2.1$, and photometric data, described in \S~$2.2$. All targets, coordinates, and photometry are listed in Table 2. The locations of the targets in the observed colour-magnitude diagram ($V-(v-y)$) of the cluster are shown in the right-hand panel of Fig.\,\ref{fig:pic5}. 
 
\subsection{High and medium-high resolution spectroscopy}
All our targets have been selected from the Str\"omgren $uvby$ photometric survey carried out by F.\,Grundahl, matching the $(b-y)$ and $ c_1$ ranges spanned by the field stars analysed by \citet{Charbonnel05a}. 
In total, 349 stars were observed across the colour-magnitude diagram of NGC~6397, from just below the cluster TO point ($V\approx17$) to the end of the red giant branch (RGB) ($V\approx11.5$). Each evolutionary phase is well sampled with around 180 stars at the TO, 80 on the SGB, and 90 stars on the RGB. Here we present data from two observing runs with FLAMES on the VLT-UT2 \citep{Pasquini02}, one that collected the Li data for the whole sample in June 2006 and a second run that completed the data set with Na abundance indicators for a subset of the targets in Aug.\ 2008. The dates, set-ups, exposure times, and atmospheric conditions are summarised in Table \ref{tab:setup}.

Both sets of observations have used the GIRAFFE$+$UVES combined mode of FLAMES, with Medusa fibres allocated to the faintest targets and UVES fibres allocated to the brightest end of the RGB. During the first observing campaign (2006), we selected the high-resolution GIRAFFE set-ups H679.7 (also known as HR15) and H627.3 (HR13), which yield respectively $R=19\,000$ and 22\,300, and spectral ranges of 660.7 -- 696.5\,nm and 612.0 -- 640.5\,nm. Both settings were observed in combination with the UVES Red\,580 setting ($R=47\,000$, spectral range: 480 -- 680\,nm, with a small gap in the middle). The first GIRAFFE set-up covers the $^7$Li resonance line at 670.7\,nm, whereas the second setting covers the Na\,I doublet at 615.4/616.0\,nm,. 

Unfortunately, the detection of the Na\,I doublet at 615.4/616.0\,nm proved to be realistic only for the brightest targets of our sample, therefore preventing conclusions on the existence of a Li-Na anti-correlation. To overcome this problem, we proposed to observe a stronger Na\,I doublet (at 818.3\,nm and 819.4\,nm,) and were granted 3 hours through the Director General Discretionary Time channel in Aug.\ 2008. Because of the limited time request we carefully selected a sub-set of our initial sample, mainly including the brightest TO and SGB stars (in total, 117 stars), with both high and low Li. For these observations, we used the high-resolution GIRAFFE set-up H805.3A (also known as HR19A, characterised by $R=14\,000$ and spectral range 774.5--833.5\,nm,) in combination with the UVES Red\,860 set-up ($R=47\,000$, spectral range 760 -- 1000\,nm). 

Standard data reduction of the 2006 GIRAFFE observations was performed with the Geneva Base Line Data Reduction Software (girBLDRS), version 1.13.1. To correct for the highly elevated dark current in the upper corner of the old CCD a carefully scaled and smoothed 2D dark-frame was subtracted from each science frame. The 2008 data were obtained after the GIRAFFE CCD upgrade that took place in Apr.\,/May 2008 and contain negligible dark current. These data were reduced with the ESO GIRAFFE pipeline, version 2.\,5.\,3. UVES data reductions were performed with the FLAMES UVES pipeline, version 2.\,9.\,7. 

After initial reduction followed sky-subtraction and radial-velocity correction. Through radial-velocity measurements we identified six non-members. Furthermore, we disregarded one star that appeared to be too metal-rich to be a cluster member, three stars that clearly fall off the cluster sequence, and yet three more stars for which the observed spectra are of too poor quality. All stars that were rejected are listed with comments in Table 2. The remaining sample consists of 32 stars with spectra obtained with the UVES Red 580 set-up (we hereafter refer to this set as the 'UVES sample') and 284 stars with spectra obtained with the GIRAFFE H679.7 and H627.3 set-ups. The UVES sample and the GIRAFFE sample have 11 stars in common. A sub-sample of eight stars were observed in the UVES Red 860 set-up and 117 with the GIRAFFE H805.3A set-up. The UVES sample and GIRAFFE sample have two stars in common.

The mean barycentric radial velocity obtained for all cluster members is $18.5\pm0.2\rm\,km\,s^{-1}$, in good agreement with \citet{Lind08} ($18.1\pm0.3\rm\,km\,s^{-1}$) and \citet{Milone06} ($18.36\pm 0.09 (\pm 0.10)\rm\,km\,s^{-1}$).    

\subsection{Str\"{o}mgren photometry}
The Str\"omgren $uvby$ photometry employed for this investigation was obtained at the Danish 1.54m telescope on La Silla, Chile, and is identical to the data set used by Korn et al. (2007) and Lind et al.\ (2008). We reiterate the main points here. All $uvby$ observations were collected during a two week run in May 1997, with a large number of observations of standard stars observed on the photometric nights. The field covered is roughly 9\arcmin\ in diameter and slightly west of the cluster centre. For the photometric reductions we used the same programs and procedures as described in Grundahl et al.\ (1998, 1999, 2000, 2002a). The photometric zero points in $vby$ have errors of approximately $0\fm01$.
\nocite{Grundahl98}
\nocite{Grundahl99}
\nocite{Grundahl00}
\nocite{Grundahl02a}

The observed $(b-y)$, $(v-y)$, and $c_1$ colours were corrected for the reddening of NGC~6397, using the relations $E_{b-y}=0.74\times E_{B-V}$, $E_{v-y}=1.70\times E_{b-y}$, and $E_{c_1}=0.20\times E_{b-y}$ \citep{Crawford75}. We adopted a value of 0.179 for $E_{B-V}$, following \citet{AnthonyTwarog00}, and 12.57 for the cluster distance modulus. The observed colours of our targets show a spread around the cluster sequence, of the order of 2.2\% for $(v-y)$ and larger, 3.7\%, for $(b-y)$. To reduce the uncertainty in the photometric colours we constructed colour-magnitude fiducials for the cluster, as described in \citet{Korn07} and Lind et al.\ (2008). The observed colour is interpolated at constant $V$-magnitude onto the fiducial sequence. The main advantage of using this method is that relative effective temperatures are more precisely determined, which in turn lowers the spread in chemical abundances.

\begin{table}
      \caption{FLAMES observations. }
         \label{tab:setup}
         \centering
         \begin{tabular}{lllll}
                \hline\hline
                Date         &   UVES      & GIRAFFE  &  Exposure            & Average \\
                             &   setting   & setting  &  time [s]            & seeing [$\arcsec$] \\
                                \hline
                2006 Jun 19   &  Red 580    & H627.3      &  $2\times2775$       & 1.08  \\
                2006 Jun 19   &  Red 580    & H679.7      &  $1\times2775$       & 1.08  \\
                2006 Jun 20   &  Red 580    & H679.7      &  $2\times2775$       & 1.08  \\
                2006 Jun 20   &  Red 580    & H627.3      &  $2\times2775$       & 1.01  \\
                2006 Jun 21   &  Red 580    & H627.3      &  $2\times2775$       & 0.75  \\
                2006 Jun 21   &  Red 580    & H679.7      &  $3\times2775$       & 0.78  \\
                2006 Jun 26   &  Red 580    & H627.3      &  $3\times2775$       & 1.16  \\
                2006 Jun 26   &  Red 580    & H679.7      &  $2\times2775$       & 0.76  \\
                2006 Jun 27   &  Red 580    & H679.7      &  $1\times2775$       & 0.56  \\
                2006 Jun 27   &  Red 580    & H627.3      &  $3\times2775$       & 0.46  \\
                2006 Jun 28   &  Red 580    & H679.7      &  $3\times2775$       & 0.69  \\
                2008 Aug  3   &  Red 860    & H805.3A      &  $2\times2775$       & 0.94  \\
                2008 Aug 28   &  Red 860    & H805.3A      &  $1\times2775$       & 1.17  \\
                   \hline
         \end{tabular}
\end{table}


\begin{figure}
	\centering
		\includegraphics[angle=90,width=8cm,viewport=1cm 3cm 17cm 25cm]{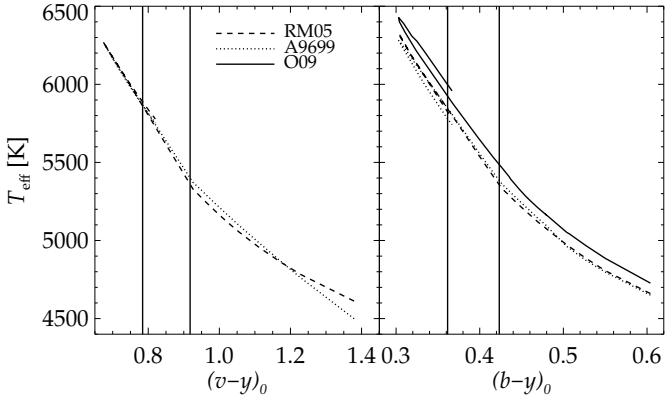}
	  \caption{The $(b-y)-T_{\rm eff}$ and $(v-y)-T_{\rm eff}$ relations considered in the analysis. The vertical lines mark the area that is linearly interpolated (see text). The O09 scale for pre-TO stars fall above the one for post-TO stars and vice versa for A9699.}
	\label{fig:pic1}
\end{figure}

\begin{figure}
	\centering
		\includegraphics[width=4.5cm,angle=90,viewport=0cm 1cm 12cm 25cm]{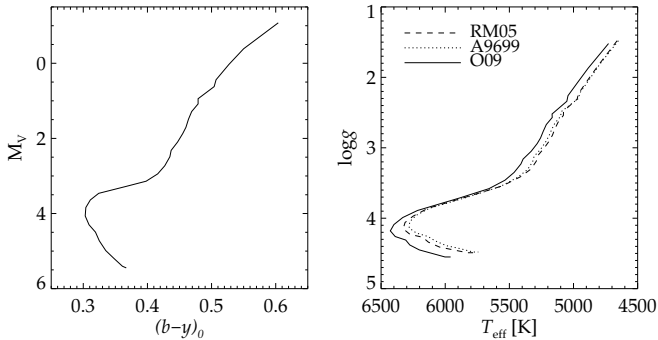}
	  \caption{\textit{Left:} The $(b-y)_0-M_V$ fiducial used to derive stellar parameters. \textit{Right:} The three $(b-y)$-based stellar-parameter sets.}
	\label{fig:pic2}
\end{figure}

\begin{table*}
\caption{Photometry and effective temperatures. The full table can be retrieved from CDS/Vizier.}
\label{tab:photo}
\centering
\tiny{
\begin{tabular}{rrrrrrrrrrrr}
\hline\hline
ID  &   RA (J2000) & DEC (J2000) &$v_{r}$           & $V$ & $(b-y)_0$ &  $(v-y)_0$ & $c_0$&  $T_{\rm eff}\rm\,O09$& $T_{\rm eff}\rm\,RM05$  & $T_{\rm eff}\rm\,A9699$ & $T_{\rm eff}\rm\,H\alpha$ \\
    &              &             &$\rm[km\,s^{-1}]$ &     &           &            &      &  [K] & [K]   &      [K]        & [K]    \\
\hline
12138  & 17  40 17.23& -53  41 57.37&  13.5 & 11.493&  0.604&  1.379&  0.452& 4727& 4660& 4650& 4540\\
16405  & 17  40 28.64& -53  38 32.42&  13.5 & 11.771&  0.582&  1.319&  0.429& 4790& 4719& 4709& 4564\\
14565  & 17  40 24.07& -53  42 52.74&  17.6 & 11.861&  0.575&  1.300&  0.422& 4811& 4738& 4728& 4569\\
5644   & 17  39 52.51& -53  39 11.24&  16.6 & 11.924&  0.570&  1.286&  0.417& 4826& 4751& 4742& 4576\\
...   & ...   &...   &...   &...   &...   &...   &...   &...   &...   &...   \\
\hline
\end{tabular}
}
\end{table*}

\begin{table}
\caption{Iron lines measured in the UVES spectrum of the RGB star \#17691. The full table can be retrieved from CDS/Vizier.}
\label{tab:iron}
\centering
\footnotesize{
\begin{tabular}{rrrrrr}
\hline\hline
Ion  & $\lambda$ [nm]& $\log{gf}$ &Ref. & $W_{\lambda}$ [pm]& A(Fe)\\ 
\hline
FeI &491.9&-0.34& 1 &$ 9.43\pm0.12$& 5.38$\pm$0.03\\
FeI &498.6&-1.33& 1 &$ 4.76\pm0.05$& 5.38$\pm$0.01\\
FeI &499.4&-2.97& 1 &$ 8.24\pm0.10$& 5.55$\pm$0.03\\
FeI &501.5&-0.30& 1 &$ 4.00\pm0.07$& 5.41$\pm$0.01\\
FeI &502.8&-1.12& 1 &$ 1.97\pm0.05$& 5.36$\pm$0.01\\
FeI &504.4&-2.02& 1 &$ 1.82\pm0.08$& 5.39$\pm$0.02\\
FeI &506.0&-5.43& 1 &$ 2.22\pm0.09$& 5.64$\pm$0.02\\
FeI &508.3&-2.84& 1 &$ 8.52\pm0.09$& 5.53$\pm$0.02\\
FeI &512.7&-3.25& 1 &$ 7.19\pm0.05$& 5.56$\pm$0.01\\
FeI &514.2&-2.24& 1 &$ 3.35\pm0.06$& 5.49$\pm$0.01\\
FeI &519.9&-2.09& 1 &$ 5.19\pm0.11$& 5.47$\pm$0.02\\
FeI &521.7&-1.16& 1 &$ 3.95\pm0.08$& 5.42$\pm$0.02\\
FeI &522.6&-4.76& 1 &$ 4.71\pm0.09$& 5.60$\pm$0.02\\
FeI &523.3&-0.06& 1 &$10.25\pm0.10$& 5.30$\pm$0.02\\
FeI &524.2&-0.97& 1 &$ 2.54\pm0.05$& 5.41$\pm$0.01\\
FeI &524.7&-4.97& 1 &$ 4.03\pm0.09$& 5.66$\pm$0.02\\
FeI &530.7&-2.91& 1 &$ 4.46\pm0.10$& 5.44$\pm$0.02\\
FeI &534.0&-0.72& 1 &$ 5.98\pm0.09$& 5.43$\pm$0.02\\
FeI &536.5& 0.23& 1 &$ 3.48\pm0.05$& 5.32$\pm$0.01\\
FeI &536.5&-1.02& 1 &$ 1.95\pm0.09$& 5.23$\pm$0.03\\
FeI &538.3& 0.64& 1 &$ 5.55\pm0.04$& 5.16$\pm$0.01\\
FeI &539.7&-1.98& 1 &$12.90\pm0.09$& 5.58$\pm$0.02\\
FeI &540.6&-1.85& 1 &$12.87\pm0.07$& 5.53$\pm$0.01\\
FeI &541.5& 0.64& 1 &$ 4.91\pm0.08$& 5.12$\pm$0.02\\
FeI &550.1&-3.05& 1 &$ 8.64\pm0.08$& 5.70$\pm$0.02\\
FeI &550.7&-2.79& 1 &$ 9.46\pm0.08$& 5.68$\pm$0.02\\
FeI &556.7&-2.67& 1 &$ 0.91\pm0.03$& 5.39$\pm$0.02\\
FeI &570.2&-2.14& 1 &$ 2.79\pm0.07$& 5.40$\pm$0.02\\
FeI &595.7&-4.50& 1 &$ 1.66\pm0.06$& 5.49$\pm$0.02\\
FeI &606.5&-1.41& 1 &$ 5.94\pm0.11$& 5.32$\pm$0.02\\
FeI &613.7&-1.41& 1 &$ 7.90\pm0.12$& 5.55$\pm$0.03\\
FeI &615.2&-3.37& 1 &$ 0.84\pm0.07$& 5.53$\pm$0.04\\
FeI &621.3&-2.48& 1 &$ 3.24\pm0.07$& 5.42$\pm$0.01\\
FeI &621.9&-2.45& 1 &$ 4.08\pm0.09$& 5.52$\pm$0.02\\
FeI &624.1&-3.17& 1 &$ 0.94\pm0.03$& 5.43$\pm$0.02\\
FeI &625.3&-1.77& 1 &$ 6.90\pm0.07$& 5.63$\pm$0.01\\
FeI &626.5&-2.54& 1 &$ 3.66\pm0.08$& 5.51$\pm$0.02\\
FeI &634.4&-2.88& 1 &$ 1.03\pm0.10$& 5.42$\pm$0.05\\
FeI &639.4&-1.58& 1 &$ 7.47\pm0.08$& 5.58$\pm$0.02\\
FeI &641.2&-0.72& 1 &$ 4.02\pm0.06$& 5.43$\pm$0.01\\
FeI &643.1&-1.95& 1 &$ 6.75\pm0.07$& 5.49$\pm$0.01\\
FeI &648.2&-3.01& 1 &$ 1.42\pm0.12$& 5.53$\pm$0.04\\
FeI &649.5&-1.24& 1 &$ 8.87\pm0.08$& 5.50$\pm$0.02\\
FeI &649.9&-4.69& 1 &$ 1.28\pm0.06$& 5.63$\pm$0.02\\
FeI &660.9&-2.66& 1 &$ 1.34\pm0.08$& 5.46$\pm$0.03\\
FeI &666.3&-2.46& 1 &$ 2.72\pm0.09$& 5.50$\pm$0.02\\
FeI &667.8&-1.42& 1 &$ 6.65\pm0.15$& 5.53$\pm$0.03\\
FeI &675.0&-2.58& 1 &$ 2.12\pm0.11$& 5.47$\pm$0.03\\
\multicolumn{6}{c}{       $<A\rm(Fe)_{I}>=5.45\pm0.02$}\\
\hline
FeII&492.4&-1.32&  4 &$10.30\pm0.13$& 5.37$\pm$0.03\\
FeII&510.1&-4.14&  2 &$ 0.34\pm0.07$& 5.38$\pm$0.09\\
FeII&519.8&-2.23&  2 &$ 4.87\pm0.10$& 5.45$\pm$0.02\\
FeII&523.5&-2.15&  2 &$ 5.13\pm0.13$& 5.40$\pm$0.03\\
FeII&526.5&-3.13&  3 &$ 1.25\pm0.15$& 5.46$\pm$0.06\\
FeII&528.4&-3.19&  3 &$ 2.48\pm0.10$& 5.52$\pm$0.02\\
FeII&532.6&-3.22&  2 &$ 1.21\pm0.10$& 5.52$\pm$0.04\\
FeII&541.4&-3.64&  3 &$ 0.54\pm0.08$& 5.55$\pm$0.07\\
FeII&542.5&-3.37&  2 &$ 1.02\pm0.10$& 5.55$\pm$0.05\\
FeII&553.5&-2.87&  3 &$ 2.35\pm0.03$& 5.54$\pm$0.01\\
FeII&614.9&-2.72&  2 &$ 0.74\pm0.08$& 5.49$\pm$0.05\\
FeII&624.8&-2.33&  2 &$ 1.64\pm0.10$& 5.49$\pm$0.03\\
FeII&636.9&-4.25&  2 &$ 0.46\pm0.07$& 5.68$\pm$0.07\\
FeII&645.6&-2.08&  2 &$ 2.37\pm0.09$& 5.45$\pm$0.02\\
\multicolumn{6}{c}{       $<A\rm(Fe)_{II}>=5.49\pm0.02$}\\
\hline
\multicolumn{6}{l}{(1) \citet{Obrian91} \ \ \ \ \ \ \ \ \ \ \ (2) \citet{Biemont91}} \\
\multicolumn{6}{l}{(3) \citet{Raassen98} \ \       (4) \citet{Fuhr88}   }\\
\end{tabular}
}
\end{table}

\section{Analysis}

For the abundance analysis we used a grid of 1D, LTE, plane-parallel and spherical, MARCS model atmospheres \citep{Gustafsson08}. LTE abundances of Li and Na were derived from spectrum synthesis, using the Uppsala code \textsc{bsyn}. Non-LTE abundance corrections were thereafter applied. LTE abundances of Fe and Ca were derived from equivalent width measurements, using the corresponding Uppsala code \textsc{eqwidth}.   

\subsection{Effective temperatures and surface gravities}

The cluster $uvby$-photometry was used to calculate effective temperatures for the whole sample. We implemented a number of different calibrations based on the colour indices $(v-y)$ and $(b-y)$ and $c_1$ . Table \ref{tab:photo} lists effective temperatures calculated using the relations published in \citet{Alonso96} and \citet{Alonso99} (hereafter these two papers are referred to as A9699), \citet{Ramirez05} (hereafter RM05), and \citet{Onehag09} (hereafter O09). The first three are calibrated on the infra-red flux method (IRFM), whereas the last is based on theoretical colours from MARCS model atmospheres \citep{Gustafsson08}. For all stars we assumed $\rm [Fe/H]=-2.0$. The IRFM colour-$T_{\rm eff}$ relations constructed for main sequence (MS) stars were calibrated on stars with $\log{g}\ga3.8$. We accordingly set a limit in absolute visual magnitude, $M_V>3.3$, for which the MS calibrations are trusted. Analogously, the giant calibrations were calibrated on stars with $\log{g}\la3.5$ and we trust them for $M_V<2.8$. This leaves a gap on the SGB where no calibration is suitable. In this range we interpolated linearly between each dwarf- and giant-calibration pair (see vertical lines in Fig.\ \ref{fig:pic1}). To each of the obtained effective-temperature scales we calculated surface gravities from the relation between surface gravity, effective temperature, stellar mass, and luminosity. The details of the calculations are given in Lind et al.\ (2008).  

O09 have calculated synthetic $(b-y)$-colours for a grid of effective temperatures, surface gravities, and metallicities. We interpolated in the grid constructed for $\rm[Fe/H]=-2.0$, which spans the parameter-space of the sample (see Fig. 4 in O09). The interpolation was done iteratively until the parameters found for each star were fully consistent, i.e.\ the $T_{\rm eff}-\log{g}$ pair found based on the observed colour are also constrained by relation to mass and luminosity. 

For the 32 RGB stars observed with UVES, we could also derive H$\alpha$-based effective temperatures, using \textsc{bsyn}. Stark broadening is based on the tabulations by \citet{Stehle99} and self-broadening of hydrogen follow \citet{Barklem00}. The fitting method is automated and based on metal line-free regions extending up to $\pm5$\,nm from the line centre \citep[see][]{Lind08}, avoiding the line core. The wings of the H$\alpha$-line become very narrow high up on the RGB, but some sensitivity to effective temperature remains also for these stars. With a high $S/N$-ratio it is possible to constrain $T_{\rm eff}$ also for the coolest stars in our sample.   
 
Figure \ref{fig:pic2} shows a comparison between all $(b-y)-T_{\rm eff}$ and $(v-y)-T_{\rm eff}$ relations. The O09 $(b-y)$-calibration matches the slope of the corresponding IRFM-based calibrations well, but is offset to higher effective temperatures by approximately 100\,K. However, a recent updated calibration based on the IRFM finds effective temperatures that are higher than the existing IRFM calibrations, by similar order of magnitude, at these metallicities (Casagrande et al. 2009, in preparation). Interestingly, the O09 scale predicts slightly higher effective temperatures for dwarfs compared to SGB stars of the same colour, whereas the IRFM of A9699 has the opposite behaviour (through the sensitivity to the $c_1$-index). The interpolation between dwarf and giant calibrations appears more suitable for the $(b-y)$-relations. The two $(v-y)$ MS scales are too flat to match the slope of the corresponding giant relations well. For this reason we regard the $(b-y)$-relations as more trustworthy and place greater weight on those. 
The H$\alpha$-based values are generally cooler than all photometric calibrations, but the relative agreement is satisfactory. 

The right panel of Fig.\ \ref{fig:pic2} shows all $(b-y)$-based effective-temperature scales with corresponding surface gravity values. Apparently, the two IRFM scales agree well for all but the hottest stars, where the A9699 scale predicts cooler effective temperatures. Again it is seen that the theoretical MARCS scale is offset to higher effective temperatures. The left panel of Fig \ref{fig:pic2} shows the cluster fiducial for $(b-y)_0-M_V$.

\begin{figure}
	\centering
		\includegraphics[width=4.5cm,angle=90,viewport=0cm 1cm 12cm 25cm]{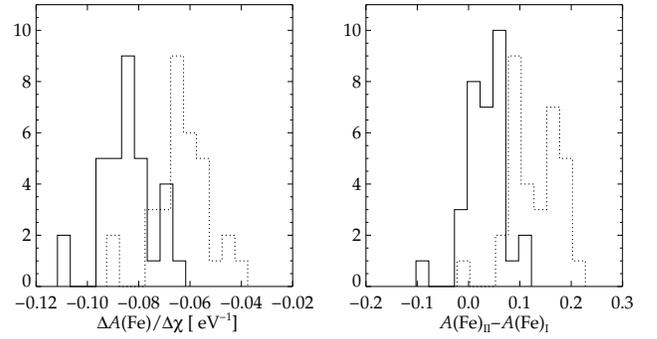}
	  \caption{\textit{Left:} Histogram of the slope of Fe abundance with excitation potential of neutral and singly ionised Fe lines. Only the UVES sample is shown. Solid lines represent the O09 effective temperature scale and dotted lines the A9699 scale. \textit{Right:} Histogram of Fe abundance derived from Fe\,II lines minus the Fe abundance derived from Fe\,I lines. }
	\label{fig:pic3}
\end{figure}

\subsection{Metallicity}
To constrain the cluster metallicity and microturbulence values, we calculated iron abundances from Fe\,I and Fe\,II lines for the UVES sample. The ionisation and excitation equilibrium of Fe also provide an additional check of the derived stellar parameters. We selected 48 unblended Fe\,I lines that have oscillator strengths determined by \citet{Obrian91} and 14 Fe\,II lines that have oscillator strengths determined by \citet{Biemont91}, \citet{Raassen98}, or \citet{Fuhr88}. Equivalent widths were measured by fitting Gaussian line profiles to the observed, fixing the FWHM of lines weaker than 5\,pm to the average value for each individual star. In the GIRAFFE spectra only very few Fe\,I and Fe\,II lines are available and it is not possible to obtain reliable measurements for the fainter half of the sample. Table \ref{tab:iron} lists all Fe\,I and Fe\,II lines with references to the adopted $\log{gf}$ values, as well as the equivalent widths and abundances derived for one RGB star. The full table, including all UVES targets, can be retrieved from the CDS. The table lists also the average abundance derived from Fe\,I and Fe\,II lines, respectively, with the error bar representing the standard deviation of the mean abundance.

Microturbulence ($\chi_{\rm t}$) values were determined by requiring that Fe abundances show no trend with reduced equivalent width ($\log(W_\lambda/\lambda)$) of Fe\,I lines. We empirically constructed a linear parameterisation of microturbulence with surface gravity, using the UVES targets only, and adopted it for the whole sample. This involved extrapolation to the SGB stars and dwarfs, since our observations do not cover enough strong lines that are sensitive to microturbulence in these stars. The impact on the inferred Li abundances is however negligible. The largest equivalent width measured for the Li line is 6.18\,pm, for the SGB star \#13160. Varying the microturbulence between $1\rm\,km\,s^{-1}$ and $2\rm\,km\,s^{-1}$ lowers the Li abundance inferred for this star by only 0.01\,dex.
  
The excitation equilibrium of Fe is commonly used to constrain effective temperatures. However, 3D modelling of stellar atmospheres show that especially low excitation lines may give strongly overestimated abundances in 1D models of metal-poor stars \citep{Asplund99,Collet07}. In addition, non-LTE effects are likely to have different impact on lines of different excitation potential \citep[e.g.][]{Asplund05}. The extent to which this could bias the excitation equilibrium is presently not clear. The left-hand histogram in Fig.\ \ref{fig:pic3} shows the slope of Fe abundances with excitation potential of Fe\,I lines, derived for the A9699 and O09 scales. Apparently, both scales are too hot to satisfy the equilibrium. The necessary corrections are of order $(-100)-(-200)$\,K. 

The right-hand histogram in Fig \ref{fig:pic3} shows the Fe abundance derived for singly ionised Fe lines minus the Fe abundance derived from neutral Fe lines. The A9699 scale shows an offset from zero of approximately $-0.14$\,dex, whereas the O09 scale has a smaller offset, $-0.05$\,dex. The mean Fe abundance, as determined from Fe\,II lines, is $5.41\pm0.01$, using both scales, with a $1\sigma$ dispersion of 0.04\,dex. Adopting a solar abundance $A\rm(Fe)_{\sun}=7.51$ (Asplund et al.\ 2009 in prep.\,), we obtain [Fe/H]$=-2.10$. This is in good agreement with most recent estimates made for RGB stars in this cluster; Korn et al (2007) find $-2.12\pm0.03$ for their RGB sample, \citet{Gratton01} [Fe/H]$=-2.03\pm0.02\pm0.04$, and \citet{Castilho00} [Fe/H]$=-2.0\pm0.05$. Lind et al.\ (2008) report a lower value, [Fe/H]$\approx-2.3$, for stars at the base of RGB. The last study implements a cooler effective temperature-scale and base the abundances mainly on Fe\,I lines, which explains the offset.  

The investigation is not indicating that one of the stellar-parameter sets is clearly preferable to the other, but we used the O09 scale to derive abundances of all other elements, because this is the most homogeneous. The changes in Li abundances when implementing the IRFM stellar-parameter scales are discussed in \S~$5.2$.

\begin{figure*}
	\centering
		\includegraphics[width=16cm,angle=90]{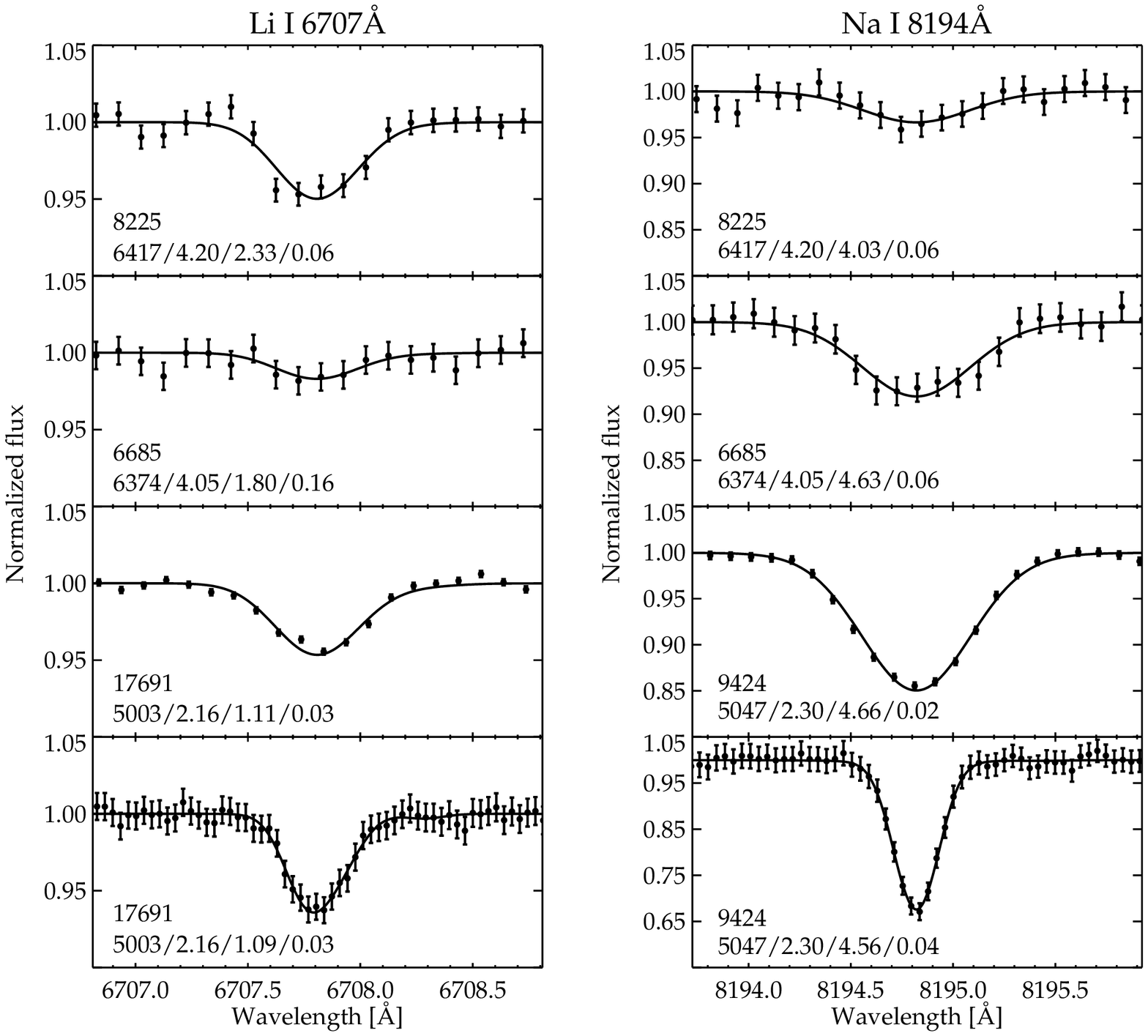}
	  \caption{Example fits of the Li\,I 670.7\,nm line and the Na\,I 819.4\,nm line. In each panel is indicated the identification number of the star, stellar parameters, and LTE abundance inferred from the given line, according to the syntax $T_{\rm eff}/\log{g}/A\rm(x)/\it e\rm(x)$, where x is the corresponding element and $e(x)$ the error in abundance. \textit{Top:} A Li and Na-normal TO star. \textit{Second from the top:} A Li-poor and Na-rich TO star. \textit{Second from the bottom:} GIRAFFE giant spectrum. \textit{Bottom:} UVES spectrum of same giant.} 
	\label{fig:pic4}
\end{figure*}

\subsection{Lithium}
The $^7$Li resonance line at 670.7\,nm has two fine-structure components, separated only by 0.015\,nm. At the resolution of GIRAFFE ($R=19300$) they are indistinguishable, but a small line-asymmetry can be seen in the higher-resolution UVES spectra ($R=47000$). Accurate measurements of the line require spectrum synthesis (see example fits of the line in Fig.\ref{fig:pic4}). We adopted $\log{gf}=-0.009$ and $-0.309$ for the two components respectively, according to \citet{Lindgard77}. 

To derive Li LTE abundances for the stars we first normalised each spectrum by iterative fitting of a first order polynomial, taking into consideration a wavelength region extending $\pm 0.2$\,nm from the line centre. Synthetic line profiles were convolved with a Gaussian profile to account for the instrumental broadening and broadening due to rotation and macroturbulence. The width of the convolving profile was fixed for each star by measurements of single lines of other species. The Li abundance was then varied to find the best fitting synthetic profile for each star through $\chi^2$-minimisation. In total, Li abundances or upper abundance limits could be obtained for 305 stars. Following \citet{Norris01} and \citet{Barklem05}, errors due to observational uncertainties were estimated with the expression: $\sigma _{W_\lambda}=\lambda\sqrt(n)/(R\times S/N)$, where $n$ denotes the number of pixels spanning the full-width at half maximum for the line, $R$ the spectral resolution, and $S/N$ the signal-to-noise ratio per pixel. 

Non-LTE abundance corrections for Li were calculated according to \citet{Lind09a}, a study that is based on calculations on the same MARCS model atmospheres as here \citep{Gustafsson08}. In contrast to earlier non-LTE analyses \citep[e.g.][]{Carlsson94,Takeda05} the study makes use of rigorous quantum mechanical calculations (as opposed to the debated but common classical recipes) of cross-sections for collisions with neutral hydrogen, which have influence over the statistical equilibrium of Li. Lind et al.\, (2009) show that including  the charge transfer reactions between Li and hydrogen in the non-LTE calculations gives abundance corrections that are lower by almost $-0.1$\,dex for dwarfs at [Fe/H]$=-2.0$, as compared to neglecting them. For our sample, the MS, TO, and early SGB stars all have non-LTE corrections that are very similar, around $-0.06$\,dex. With decreasing effective temperature and surface gravity the corrections change sign and reach a maximum of $+0.13$\,dex for the coolest RGB star.

\subsection{Sodium} 
Sodium abundances or upper abundance limits were derived from the 2008 observations for a subset of 117 stars, mainly TO stars and early SGB stars. The abundances were based on the Na\,I 818.3\,nm and 819.4\,nm doublet (when the former was too weak to be detected, only the latter was used). This doublet is second in strength after the resonance doublet at 588.5\,nm and 589.0\,nm, which is strongly affected by interstellar extinction for the cluster stars and is therefore not useful as abundance indicator. As mentioned above, the 568.2/568.8\,nm and 615.4/616.0\,nm doublets are too weak to be suitable for abundance analysis in the TO region at this metallicity. Despite its unfortunate location in the middle of a strong telluric band, the 818.3/819.4\,nm Na\,I doublet does not suffer strongly from atmospheric blends that would bias the abundance analysis.

The line-fitting method is the same as described for Li and oscillator strengths by \citet{Kurucz75} were adopted. We apply non-LTE abundance corrections to the 818.3/819.4\,nm Na\,I lines according to \citet{Mashonkina00}. The size of the corrections at this metallicity ranges from $-0.14$\,dex for the hottest, highest surface gravity stars, to approximately $-0.31$\,dex for the coolest giants.

\begin{table*}
      \caption{Adopted stellar parameters, equivalent widths, and Li, Na, and Ca abundances. The full table can be retrieved from CDS/Vizier.}
         \label{tab:abund}
         \centering
\tiny{
   \begin{tabular}{p{0.55cm}p{0.8cm}p{0.35cm}p{0.35cm}rrp{1.0cm}p{1.0cm}p{1.0cm}p{1.0cm}p{1.0cm}p{1.0cm}p{1.0cm}p{1.0cm}}
\hline\hline
ID & $M_V$ &$T_{\rm eff}$ & $\log{g}$ & $\xi$ & $\log{\rm L/L_{\sun}}$ &$W_{\lambda}\,670.7$  & A(Li) & $W_{\lambda}\,818.3$ & $W_{\lambda}\,819.5$ &  A(Na) & $W_{\lambda}\,612.2$ & $W_{\lambda}\,616.2$ & A(Ca) \\
   &       & [K]          &           & [km/s] & &[pm]&&[pm]&[pm]&&[pm]&[pm]&\\
\hline
12138  & -1.077&4727& 1.52& 1.61& 2.49&$ < 0.44$&$ < 0.08$&8.85 $\pm0.31$&11.49 $\pm0.17 $&4.29 $\pm0.03 $&10.91 $\pm0.08 $&12.19 $\pm0.06 $&4.70 $\pm0.01 $\\		  
16405  & -0.799&4790& 1.66& 1.58& 2.37&$ < 0.42$&$ < 0.17$&$  ...        $&$  ...        $&$ ...         $&10.41 $\pm0.06 $&11.64 $\pm0.02 $&4.64 $\pm0.01 $\\		  
14565  & -0.709&4811& 1.71& 1.58& 2.33&$ < 0.43$&$ < 0.19$&$  ...        $&$  ...        $&$ ...         $&10.30 $\pm0.02 $&11.39 $\pm0.07 $&4.62 $\pm0.01 $\\		  
5644   & -0.646&4826& 1.74& 1.57& 2.30&$ < 0.55$&$ < 0.25$&$  ...        $&$  ...        $&$ ...         $&10.25 $\pm0.08 $&11.40 $\pm0.09 $&4.62 $\pm0.01 $\\
17163  & -0.439&4873& 1.84& 1.56& 2.21&0.43  $\pm0.13 $ &0.19 $\pm0.14 $&$  ...        $&$  ...        $&$ ...     $&9.97 $\pm0.07 $&11.07 $\pm1.2 $&4.62 $\pm0.01$\\
 ... & ... & ... &...& ... & ...& ... & ... & ... & ... &...& ... & ...  \\
\hline
\end{tabular}
}
\end{table*}

\subsection{Calcium} 
For the complete sample of 305 stars, only the equivalent widths of Ca lines 612.2\,nm and 616.2\,nm could be reliably measured. We based the Ca abundance on these two lines for 305 stars, adopting oscillator strengths from \citet{Smith75}

\section{Results}

\begin{figure*}
	\centering
		\includegraphics[width=11cm, angle=90,viewport=0cm 1cm 15cm 25cm]{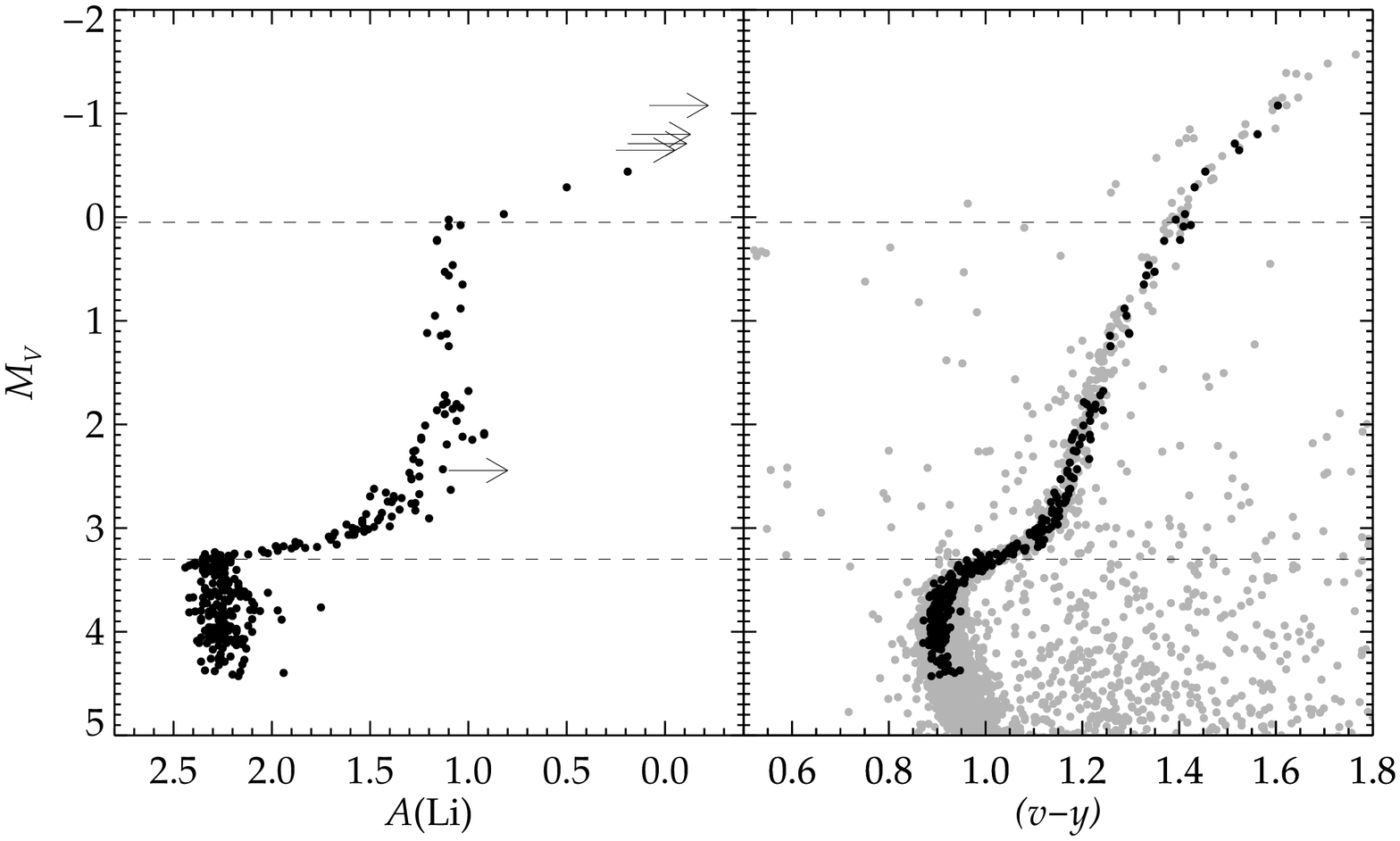}
	\caption{\textit{Left:} The abscissa shows non-LTE Li abundances inferred for our sample (the archival data are not included) and the ordinate absolute visual magnitude ($M_V=V-12.57$). Arrows mark Li upper limits. \textit{Right:} The spectroscopic targets marked with black filled circles in the colour-magnitude diagram of NGC~6397. The two horizontal dashed lines mark the location of rapid decrease in Li abundance caused by stellar evolution (see text).}
	\label{fig:pic5}
\end{figure*}

\subsection{Li abundances}
Table \ref{tab:abund} lists the adopted stellar parameters, equivalent widths, and abundances. Fig. \ref{fig:pic5} shows the Li abundances inferred side-by-side with a colour-magnitude diagram where the spectroscopic targets are marked. In agreement with the expectations (see \S~1), there is a rather well defined abundance plateau among the least evolved stars, followed by a drastic drop in Li abundance in the middle of the SGB ($M_V\approx3.3$). This pattern is caused by the so-called first dredge-up, i.e., the dilution of the external convective stellar layers with deeper hydrogen-processed material. A second steep drop occurs at $M_V\approx0.0$, which corresponds to the luminosity of the RGB bump. A similar decrease in Li abundance have been identified in stars located around the RGB bump in NGC~6752 \citep{Grundahl02b} as well as in stars in the halo field \citep{Gratton00}. \citet{Charbonnel07} describe how this second Li abundance drop can be explained by a mixing process called thermohaline convection, which becomes efficient when the hydrogen-burning shell crosses the chemical discontinuity left behind by the first dredge-up, and which rapidly transports surface Li down to internal hotter regions where this fragile element burns.

The average Li abundance on the plateau, i.e.\ only including stars that have not undergone dilution due to the first dredge-up ($M_V>3.3$), is $2.25\pm 0.01$, where the error is the standard deviation of the mean. The 1$\sigma$ dispersion is 0.09\,dex.  

\subsection{Lithium data from the ESO archive}
\begin{figure}
	\centering
		\includegraphics[width=7cm, angle=90,viewport=1cm 1cm 20cm 25cm]{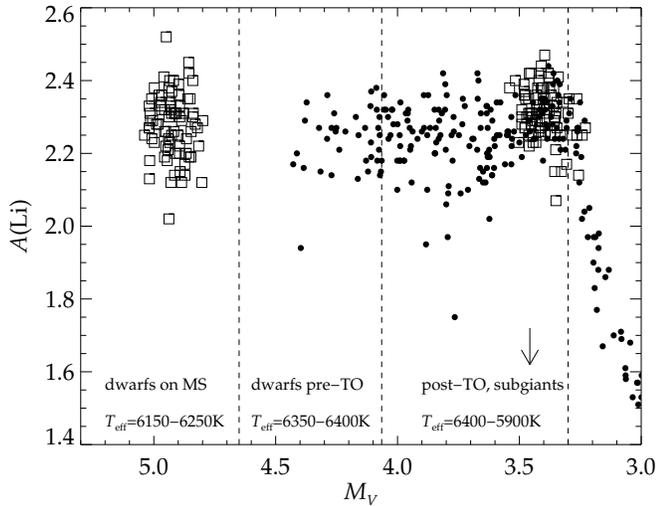}
	\caption{Lithium abundance against visual magnitude. The targets of our observations are plotted with filled circles and the archival data are plotted with open squares. Information about evolutionary status and effective temperature for various $M_V$ intervals is also given.}
	\label{fig:pic6}
\end{figure}

To establish a view of the Li content in NGC~6397 that is as complete as possible, we have searched the ESO archive for observations of more MS, TO, and SGB stars. Especially, to put constraints on the size and significance of a small increase in $A\rm(Li)$ located just before the onset of the first dredge-up (see \S~$5.1$ and Fig. \ref{fig:pic6}) we would benefit from having more stars. In previous analyses, \citet{Bonifacio02a} analyse Li in twelve TO stars (of which seven stars were previously analysed by \citeauthor{Thevenin01} 2001 and three stars by \citeauthor{Pasquini96} 1996) and \citet{Korn07} analyse five TO stars and two SGB stars. We did not incorporate those stars in our analysis since it would only slightly have increased the number statistics. However, we have retrieved and analysed a set of FLAMES archival data from 2007 (079.D-0399(A), P.\,I.\, Gonzalez-Hernandez). These include observations in the GIRAFFE high-resolution set-up H665.0 (HR15N, $R=17000$), covering the LiI 670.7\,nm line for a sample of 80 dwarfs ($M_V\approx$4.9) and 88 SGB stars ($M_V\approx3.4$) in NGC~6397. The reduction of the data was performed with the girBLDRS pipeline. In total, 38 MS stars and 55 SGB stars were identified in our $uvby$ photometric catalogue. For the remaining stars, we add a zero-point correction of 0.09\,magnitudes to the visual magnitude scale listed in the archival data, estimated from the targets that were cross-matched between the catalogues. Applying the correction, the two magnitude scales match to typically within 0.02 magnitudes for individual targets. After establishing the visual magnitudes, colours are assigned to each target by interpolating $V$ onto the fiducial cluster sequences, as was done for the main sample (see \S~$2.2$).
 
The results of the analysis of the archival data are shown together with our own data set in Fig. \ref{fig:pic6}. One more very Li-poor star, marked with an arrow, was identified in the archive sample, for which only an upper limit to Li could be derived. Especially, the region around $M_V=3.2-3.5$ is now much better sampled, with the archival data confirming the presence of the small abundance increase just before the steep drop. In addition, the plateau is extended to fainter magnitudes, $M_V\approx5$. Including this archive sample we have Li abundances for 454 stars in the cluster, of which 346 have $M_V>3.3$. Overall the quality of the data sets are similar and the abundances agree very well between the samples. In total, 14 SGB stars overlap and thus have observations of the Li line in the H665.0 as well as in the H679.7 setting. The mean difference in Li abundance between these 14 stars is $-0.02\pm0.02\,$dex, i.e. there is no systematic bias between the data sets. 

The average Li abundance on the plateau (i.e., excluding stars with $M_V<3.3$), including both our data and the archival data, is $2.270\pm0.005$. Unfortunately Na information is not available for the archive targets.

\subsection{The Li-Na anti-correlation}

As discussed in \S~1, globular-cluster samples may suffer from bias due to early pollution of the intra-cluster gas by a first generation of more massive, faster evolving stars. It is thus important to discriminate between first-generation stars that formed out of pure pristine material and second-generation stars that formed out of polluted material. The former should have been born with the cosmological Li abundance and a low Na-abundance, similar to that of field stars of comparable overall metallicity. The latter may have inherited a lower Li abundance at birth and will also exhibit other signatures of H-burning at high temperature, in particular enhanced Na abundances resulting from the NeNa chain in the polluter stars.

\begin{figure}
\centering
\includegraphics[width=7cm, angle=90]{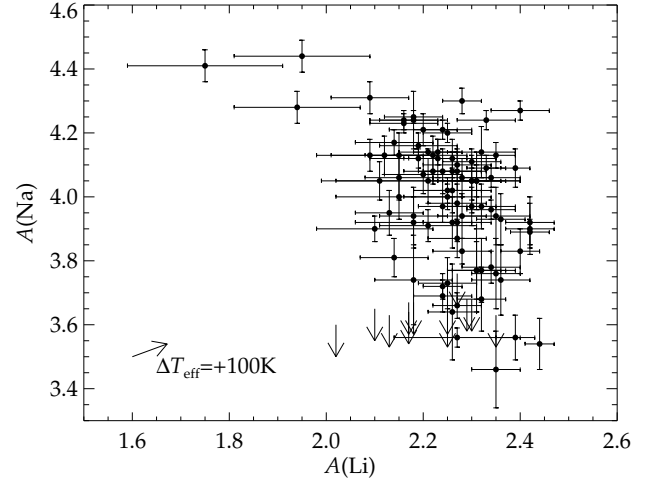}
	\caption{Non-LTE Li and Na abundances for the sub-sample of 100 dwarfs and SGB stars with observations in the H805.3A setting and $M_V>3.3$. The arrows mark upper limits to the Na abundance. If the effective temperature of a target has been underestimated by 100\,K, the corresponding point in this diagram should be shifted approximately in the direction of the tilted arrow and in its inverse direction for overestimated effective temperatures.}
	\label{fig:pic7}
\end{figure}

In Fig. \ref{fig:pic7} we report the Li and Na abundances for 100 stars in our sample with $M_V>3.3$, i.e., those that have not yet undergone Li dilution via the first dredge-up. Note that it is the first time that these quantities are determined simultaneously for such a large sample of TO and SGB stars in a globular cluster. This diagram clearly emphasises the extent of the star-to-star Na abundance variations (by up to 1~dex) within NGC~6397. This is in agreement with the findings by \citet{Carretta05}, which were based on six dwarfs and nine SGB stars. From the appearance of Fig.\,\ref{fig:pic7}, we deduce that the most Li-poor targets indeed show the most elevated Na levels, implying an anti-correlation between the abundances of Na and Li. The spectrum of the most Li-poor and second most Na-rich star in the sample, \#6685, is shown in Fig.\ \ref{fig:pic4}. The abundance pattern displayed by this star and by at least three more stars, agrees well with the globular cluster self-enrichment scenario and we conclude that they have likely been formed from polluted gas. Field stars with similar metallicity to NGC~6397 have typical Na abundances in the range $A\rm(Na)=3.6-3.9$ \citep{Andrievsky07}. It is thus reasonable to regard stars with $A\rm(Na)<3.9$ as belonging to a first generation of non-polluted objects.  

To find the statistical significance of the indicated Li-Na anti-correlation we performed linear regression between $N(\rm Li)\it /N(\rm H)$ and $N(\rm Na)/\it N(\rm H)$, taking into account measurement errors in both quantities as well as upper limits on Na \citep[IDL-routine \texttt{linmix\_err, see}][]{Kelly07}. The analysis gave a correlation coefficient of $-0.6$ between the Na and Li abundances, which for a sample of this size is highly significant. The probability of two uncorrelated variables producing such a correlation coefficient is less than 0.05\%. However, as can be realised from Fig. \ref{fig:pic7}, the significance of the anti-correlation is dependent on the most Na-enhanced stars. In fact, performing the same linear regression for stars with $A\rm(Na)<4.1$ no significant anti-correlation was found. The Li abundances among the plateau stars are thus not much affected by high degrees of pollution.  

In Fig. \ref{fig:pic7} it is also indicated with a tilted arrow approximately how the location of a star would be affected by an erroneous effective temperature. An error of $\Delta T_{\rm eff}=\pm 100$\,K corresponds to $\Delta A\rm(Li)=\pm 0.07$ and $\Delta A\rm(Na)=\pm 0.04$. Errors in effective temperature therefore weakly correlate the abundances with each other, rather than anti-correlate them. Hence, it is clear that the anti-correlation itself is not an artifact from uncertainties in the stellar-parameter determination. Possibly even a tighter anti-correlation exists for all targets, but is being distorted by the tendency of effective-temperature errors to align the abundances. As the Li abundance as well as the Na abundance are based on lines from neutral atoms, the effective temperature is the stellar parameter with by far the greatest influence over the results. The uncertainties stemming from surface gravity, microturbulence, and metallicity are all negligible in the context. 

The extent of the Li and Na variations found in the present study is similar to that found in NGC~6752 \citep{Pasquini05} and 47~Tuc \citep{Bonifacio07b}, and thus appears to be independent of the metallicity. However, the Li-Na anti-correlations in NGC~6752 and 47~Tuc are  based respectively on nine and four stars only, limiting the comparison. A more detailed discussion of the implications for the self-enrichment of NGC~6397 will be performed in a separate paper where O, Na, Mg, and Al abundance determinations will also be presented for a subset of RGB stars in our sample. In \S~5 we use the information about Na to constrain intrinsic stellar Li depletion processes.

\section{Discussion}

\subsection{Signatures of intrinsic lithium depletion}
A proposed explanation for the difference between the Li abundance found in the metal-poor halo and the BBNS prediction is intrinsic Li depletion due to the combined effect of gravitational settling and weak turbulence, the nature of which is still a matter of debate. In the case of NGC~6397, the stars on the Spite plateau would have to have been depleted by typically $\Delta A\rm(Li)=2.72-2.27=0.45$\,dex, corresponding to almost a factor of three. To constrain the physics involved in producing such depletion, it is important to accurately establish trends of Li abundance with evolutionary phase and effective temperature. In the following, we carry out our analysis based on sub-samples of data defined according to Na content. As discussed in \S~$4.3$, it is reasonable to consider that the stars born out of pristine material are those with $A\rm(Na)<3.9$.

\begin{figure}
	\centering
		\includegraphics[width=6cm, angle=90,viewport=0cm 1cm 19cm 25cm]{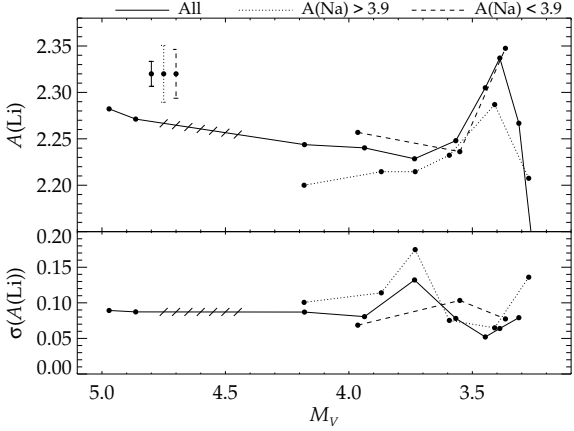}
	\caption{Group-averaged Li abundances against visual magnitude.  The full line is drawn for the whole sample (including the archival data), i.e., for both first-generation Na-poor stars and second-generation Na-rich stars. Each bin contains 40 stars. The dashed and dotted lines correspond to different sub-sets of our own sample, selected according to their corresponding Na abundances. Each bin contains 13 stars. In the upper left corner mean error bars are shown for each line. Note that the range between $M_V\approx4.4-4.8$ is not sampled by the data. }
	\label{fig:pic8}
\end{figure}

\begin{figure}
	\centering
		\includegraphics[width=6cm, angle=90,viewport=0cm 1cm 19cm 25cm]{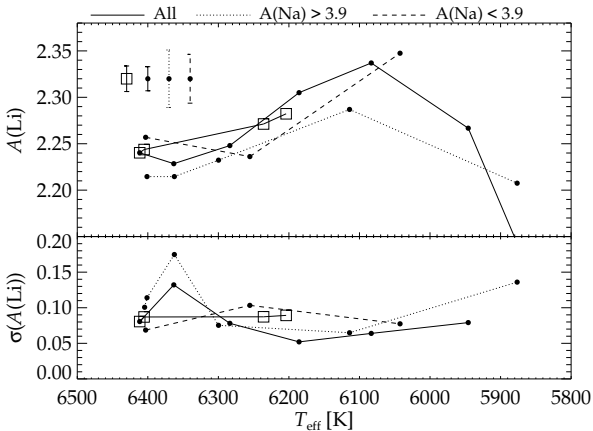}
	\caption{Group-averaged Li abundances against effective temperature. The lines correspond to the same sub-samples as in Fig.~\ref{fig:pic8}.  Filled circles represent post-TO stars and open squares pre-TO stars. In the upper left corner mean error bars are shown for each line. }
	\label{fig:pic9}
\end{figure}

Fig.~\ref{fig:pic8} shows Li abundance vs visual magnitude on the plateau, with abundances binned in groups of targets for clarity. In the lower panel of each plot, the 1$\sigma$ dispersions in the corresponding bins are shown. The full lines are drawn for the whole sample including both our data and the archival data, i.e., it does not discriminate between first-generation Na-poor stars and second-generation Na-rich objects. The single plateau-star for which only an upper limit could be inferred is not included in the binning. The dashed lines correspond to first-generation stars with $A\rm(Na)<3.9$ and dotted lines to second-generation stars with $A\rm(Na)>3.9$. Mean error bars for each line are also indicated in the figure. Fig.~\ref{fig:pic9} shows the same plot, but using effective temperature as the reference scale instead of visual magnitude. To separate pre- and post-TO stars with the same effective temperature, the dwarfs are represented by open square symbols and the SGB stars with filled circles. The approximate evolutionary status of the stars can be recovered from Fig. \ref{fig:pic6}. Stars with $M_V \geq 4$ are dwarf stars below the TO, having effective temperatures reaching from the maximum 6400\,K to 6200\,K at $M_V\approx5$. Stars with $M_V<4$ are slightly more massive objects that have passed the TO, i.e., SGB stars with $T_{\rm eff}\approx 6400-5900$\,K. At $M_V \approx 3.3$ the Li dilution process through the first dredge-up sets in. 

As can be seen in Figs~\ref{fig:pic8} and \ref{fig:pic9}, trends of Li abundance with evolutionary phase and effective temperature are not significantly different between first- and second-generation objects. The mean value of $A\rm(Li)$ appears generally to be slightly higher for the first-generation stars (dashed lines) than for the second generation (dotted lines), in agreement with what is expected in the globular cluster self-enrichment framework, but the trends are fully compatible within the error bars. We conclude that the mean Li abundance trends of the full sample are not significantly biased by intra-cluster pollution in NGC~6397. On examination of Fig~\ref{fig:pic8}, we especially note the presence of an upturn in $A\rm(Li)$ just before the steep abundance drop at $M_V\approx 3.3$, which appears rather robust against varying $A\rm(Na)$ in the sample. In a very limited magnitude-range the mean Li abundance increases by around 0.1\,dex up to $A\rm(Li)=2.35$. As mentioned previously, a difference of this size between TO and SGB stars in NGC~6397 was identified by Korn et al.\ (2007). 

The behaviour of Li abundance with $T_{\rm eff}$ is illustrated in Fig.\ \ref{fig:pic9}. There is an overall increase with decreasing effective temperature in the range $6400-6100$\,K, where a maximum is reached, followed by a decrease again until $T_{\rm eff}\approx 5900$\,K, where the first dredge-up sets in. The initial increase appears to be present both for pre- and post-TO stars. Speculatively, the lower value found for hotter dwarfs compared to cooler dwarfs, could be interpreted as the counterpart of the very right-shoulder of the so-called Li-dip observed in Population I stars \citep[see references in][]{Talon98,Talon03}. This feature appears in the present data at exactly the same effective temperature as in open clusters: In the Hyades, the mean Li abundance decreases by approximately $0.10-0.15$\,dex in the effective temperature range between 6200 and 6400~K. As discussed by \citet{Talon98,Talon04}, a similarity between Pop\,I and Pop\,II dwarfs is theoretically expected, since the depth of the stellar convective envelope, and thus the nature and the efficiency of the Li depletion mechanism, are expected to depend only on the stellar effective temperature\footnote{At a given effective temperature, Pop II stars along the plateau have lower masses than Pop I stars on the cool side of the Li dip. Regarding the depth of the convective envelope however, their lower metallicity compensates for the mass effect. See \citet{Talon04} for more details.}. If this feature is real (see \S~$5.2$), it seems that we may have discovered for the first time in Population II stars the very beginning (in terms of effective temperature) of the Li-dip, which may be more shallow at this low metallicity. Note that one should be cautious with interpretations of Li trends for dwarfs in our sample. As mentioned in \S~2.1, due to observational limitations the majority of our sample stars with Na determination are SGB stars, only a handful being relatively warm dwarfs. No Na abundance determination has thus been carried out for the coolest dwarfs below the TO. However, we have seen that the bias in Li abundance trends among SGB stars due to pollution is weak, which is likely true also for dwarfs. 

In \S~4 we concluded that the most Li-deficient stars in our sample likely belong to a second generation of stars, having experienced intra-cluster pollution. Among the first-generation stars with low Na levels, there are no stars with Li abundances lower than $A\rm(Li)=2.0$. The lower panels of Figs.\ \ref{fig:pic8} and \ref{fig:pic9} show that the abundance dispersion of the full sample is always rather low, below 0.1\,dex, except for the bin at $M_V\approx3.7$ and $T_{\rm eff}\approx6350$\,K, which contains the most Li-poor star with an abundance detection (see Fig. \ref{fig:pic6}) and therefore has a higher dispersion. Focusing on the first generation of Na-poor stars only, the typical abundance scatter is 0.09\,dex, not following any obvious trends with effective temperature or visual magnitude. Not shown in the figures is the typical measurement error in abundance, stemming from photon noise, which ranges from 0.08\,dex for the hottest stars to 0.05\,dex for the coolest plateau stars. To this measurement error should be added the propagated uncertainty in stellar parameters. Assuming that the typical star-to-star error in effective temperature, which is the most influential parameter, lies in the range $50-100$\,K, a corresponding additional spread of $0.04-0.07$\,dex in Li abundance is expected. Given these basic estimates, we conclude that the observations are compatible with zero scatter in Li abundance among the first-generation stars.

\subsection{Effects of stellar parameters and non-LTE}
We now briefly discuss the influence on the abundance trends presented in \S~5.1 from the choice of effective-temperature scale and from the non-LTE treatment. Basically, both effects have a systematic influence over the Li data. If we were to adopt the IRFM-based effective-temperature scale by RM05 (see \S~3.1) the Li abundances of post-TO stars would be systematically lowered by $-0.06$\,dex, thus aggravating further the overall discrepancy to the primordial value. The A9699 scale would act in the same direction, but also have a minor differential impact on the Li abundances since this effective-temperature scale is steeper in the range from the TO to further up the SGB. The TO star abundances would therefore be lowered by a greater amount ($-0.1$\,dex) than the cooler SGB star abundances, which in turn would enhance the abundance upturn seen in Fig. \ref{fig:pic8} and the corresponding maximum abundance seen in Fig.\ \ref{fig:pic9} at $T_{\rm eff}=6100$\,K. Nothing would thus qualitatively change in the discussion of abundance trends for post-TO stars. However, implementing either of the two IRFM-based scales would lower the abundances found for cool dwarfs more than hot dwarfs. The difference between these two groups discussed in \S~5.1, which we speculated could be a signature of the Li-dip, would hence be erased.  

As mentioned in \S~3.3, the non-LTE abundance corrections are similar, approximately $-0.06$\,dex, for all plateau stars. The LTE abundances therefore have the same relative behaviour but are offset to a slightly higher values.  

\subsection{Comparison to other studies}
Comparing the found Li abundances to other determinations made for NGC~6397, we find good agreement with Korn et al.\ (2007), who place their five TO stars with $M_V\approx3.6$ at $\rm A(Li)=2.25$ and two SGB stars with $M_V\approx3.3$ at $\rm A(Li)=2.36$. The similarity between the abundances for TO and SGB stars is partly due to the cancellation between the effects from the cooler effective-temperature scale of Korn et al.\ and the fact that they do not correct for non-LTE. For RGB stars the two effects do not cancel (non-LTE corrections are positive for RGB stars) and Korn et al.\ thus find abundances that are approximately 0.1\,dex lower than ours. The Li abundances reported by Bonifacio et al. (2002) for TO stars lie on average 0.1\,dex higher than ours. This difference can be fully traced to their effective-temperature scale, which is hotter by approximately 50\,K than the one adopted in this study, and to their use of different non-LTE corrections \citep{Carlsson94}. For these stars, the corrections by Carlsson et al.\ are smaller in absolute value than the ones found by \citet{Lind09a}. The Li abundances and abundance upper limits reported by \citet{Castilho00} for 16 RGB stars are in good agreement with ours, except for the Li detections made for giants cooler than 4900\,K, which are significantly higher than what we find.     

In the extensive literature study of field stars carried out by \citet{Charbonnel05a}, the authors concluded that the stars in their TO and SGB samples generally are more Li-rich than their dwarf sample. Especially, dwarfs that are cooler than 6000\,K show a distinct decrease in Li abundance with decreasing effective temperature. As the dwarfs in our sample are all hotter than 6000\,K we cannot verify this result. However, with hindsight one may trace a tendency of increasing Li abundance with decreasing effective temperature in the range $T_{\rm eff}=6050-6400$\,K among the MS, TO, and post-TO stars observed by Charbonnel \& Primas (see their online tables 5-9), similar to the one we find.

\begin{figure}
	\centering
		\includegraphics[width=8cm, angle=90,viewport=0cm 7cm 17cm 25cm]{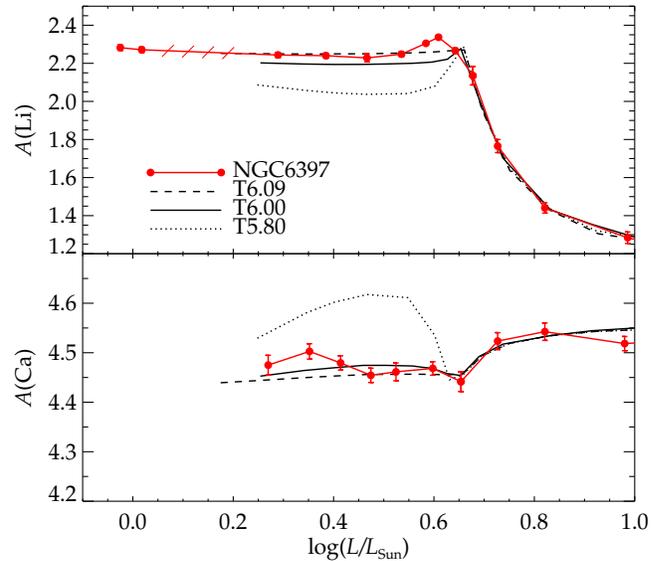}
	\caption{\textit{Top:} Comparison between bin-averaged Li abundances (red filled circles connected with solid lines) and the predictions from the stellar-structure models of Richard et al. (2005). T$5.80$ represents the model with lowest efficiency of turbulent transport, T$6.00$ intermediate efficiency, and T$6.09$ highest efficiency. The reference scale is logarithmic luminosities in units of solar luminosities. \textit{Bottom:} The same plot for Ca abundances. A colour version of this figure is available in the online edition of the journal.}
	\label{fig:pic10}
\end{figure}

\begin{figure}
	\centering
		\includegraphics[width=8cm, angle=90,viewport=0cm 7cm 17cm 25cm]{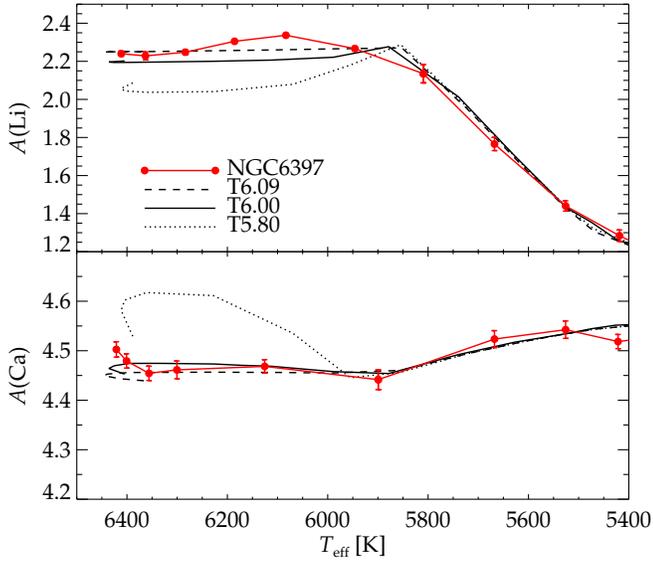}
	\caption{\textit{Top:} Same as in in Fig\ \ref{fig:pic10} but with effective temperature as the reference scale. Here, only post-TO stars are included. \textit{Bottom:} The same plot for Ca abundances. A colour version of this figure is available in the online edition of the journal.}
	\label{fig:pic11}
\end{figure}

\subsection{Comparison to diffusion-turbulence models}
Fig.\ \ref{fig:pic10} and Fig. \ref{fig:pic11} show comparisons of our obtained Li abundances with the predictions from stellar-structure models including atomic diffusion from first principles and an ad-hoc recipe for turbulent mixing \citep[and references therein]{Richard05a}. For clarity we show the same group-averaged Li abundances as in Fig.~\ref{fig:pic8} and Fig.~\ref{fig:pic9}, including the whole sample, irrespectively of Na abundance. Only post-TO stars appear in Fig.\ \ref{fig:pic11}. For comparison, the bottom panels of both figures show corresponding trends for binned Ca abundances. Visual magnitudes are converted to stellar luminosities using the bolometric corrections of \citet{Alonso99}. The same three models, T$5.80$, T$6.00$, and T$6.09$, with different efficiencies of turbulent transport, as displayed in Korn et al. (2007) and Lind et al.\ (2008) are shown. To ensure that the model $T_{\rm eff}-\log(\rm L/L_{\sun})$-plane is in agreement with our study regarding the location of the TO-point, we add +80\,K to the effective-temperature scale of Richard et al. and +0.05\,dex to the logarithmic luminosity-scale. The absolute abundance scale of the models is shifted to agree with the observations for stars evolved beyond the onset of the first dredge-up. 

The efficiency of turbulent mixing determines how much Li that is transported between the external convection zone to the region in the stellar interior where Li is destroyed, which in turn has influences over the appearance of the Li abundance plateau. In the T$6.09$ model, with highest efficiency, the plateau is flat, whereas the surface abundances predicted by the T$5.80$ model show a clear dependence on stellar luminosity and effective temperature. As seen in Fig.\ \ref{fig:pic10}, the Li abundances of NGC6397 strictly limit the efficiency of turbulent mixing, although no model perfectly reproduces the observed abundances. In terms of the size of the Li variations on the plateau, the T$6.00$ model appears to be the best choice, supporting the findings by Korn et al.\ (2007) and Lind et al.\ (2008). The observed Ca abundance trend is in good agreement with both higher-efficiency models (T6.00 and T6.09), and especially the location of the upturn at $\log(L/L_{\sun})\approx 0.65$ is matched. However, Fig. \ref{fig:pic11} shows clearly that the behaviour of the post-TO Li abundance with effective temperature is not reproduced by any of the models. Especially the location of the maximum $A(\rm Li)$ occurs at higher effective temperature than the models predict.

The initial Li abundance predicted by the models, accounting for the adjustment we make in the vertical direction, is $A(\rm Li)=2.46$, which is significantly closer, but still not in full agreement with the latest predictions from BBNS of $2.72\pm0.06$ \citep{Cyburt08}.

\section{Conclusions}
We have presented a comprehensive study of the Li content in a large sample, 454 stars in total, of MS, TO, SGB, and RGB stars in the metal-poor globular cluster NGC~6397. The cluster dwarfs and early SGB stars form a thin Li abundance plateau on the same level as the Spite plateau for field stars, whereas more evolved stars have undergone drastic Li depletion due to physical processes connected to low-mass stellar evolution. Two remarkably well-defined locations in absolute visual magnitudes, $M_V\approx3.3$ and $M_V\approx0.0$ respectively, are identified as the locations of corresponding steep Li abundance drops. They correspond respectively to the occurrence of the first dredge-up on the SGB and to the onset of thermohaline mixing at the RGB bump \citep{Charbonnel07}.

Using information of Li and Na abundance for a sub-sample of 100 dwarfs and early SGB, we have for the first time identified a significant anti-correlation between the two elements in this cluster. This is interpreted as the signature of intra-cluster pollution from a previous generation of more massive stars. The spread in Na and Li abundances of stars not having undergone Li depletion due to dredge-up, is very large, about one order-of-magnitude. However, only a handful of stars show significantly depleted Li levels and the identified anti-correlation depends critically on these objects. For Na-enhancements up to 0.7\,dex, no corresponding Li-deficiency can be detected. The average Li abundance is thus not much affected by internal pollution. These abundance patterns have consequences for for the self-enrichment scenario in NGC~6397, which will be discussed in a forth-coming paper.

By dividing our sub-sample with Na determinations into two groups, consisting of first and second generation of stars according to the degree of pollution, we illustrate how especially trends of Li abundance with effective temperature and evolutionary phase are in reasonable agreement between the two generations. The average Li abundance is typically $A\rm(Li)=2.25$ for stars located below and above the turn-off, and show a slight upturn of $\Delta A\rm(Li)\approx$0.1\,dex for stars in the middle of the subgiant branch. This difference agrees with with previous findings for the cluster as well as for field stars. We find no support for a significant Li dispersion among the first generation of stars.

The identification of a minor Li deficiency of the hottest dwarfs in our sample compared to slightly cooler dwarfs leads us to suggest that the very right-hand wing of the Li-dip may be visible also among Pop II stars. However, the signature is erased if a different effective temperature scale is adopted. 

The detailed picture we have formed of the Li abundance trends in this globular cluster can be used to constrain the physics involved in depleting Li from the photospheres of low-mass metal-poor stars, as well as the extent of the depletion. To illustrate this, we compare our observational findings to predictions from stellar-structure models including atomic diffusion, with additional turbulence below the convection zone. We find that some turbulence, in a very limited efficiency-range, is indeed required to explain observations. However, these models fail to reproduce the behaviour of Li abundance with effective temperature along the plateau, suggesting that a detailed understanding of the physics responsible for depletion is still lacking.

\begin{acknowledgements}
F.\,G.\ acknowledges support from the Danish AsteroSeismology Centre, the Carlsberg Foundation, and the Instrument Center for Danish Astronomy (IDA).  
C.\,C. and K.\,L. acknowledge financial support from the Swiss National Science
Foundation (FNS) and the french Programme National de Physique Stellaire (PNPS) of CNRS/INSU.
This research has made use of the VizieR catalogue access tool, CDS, Strasbourg, France.
\end{acknowledgements}

\bibliographystyle{aa}

\end{document}